\documentclass[showpacs,preprint,preprintnumbers,showkeys,floatfix,
superscriptaddress,amsmath,amssymb,nofootinbib,pra,aps]{revtex4}
\newcommand{\iu}{\mathrm{i}\mkern1mu}
\newcommand{\be}{\begin{equation}}
\newcommand{\ee}{\end{equation}}
\newcommand{\beq}{\begin{eqnarray}}
\newcommand{\eeq}{\end{eqnarray}}
\newcommand{\bee}{\begin{equation}}
\newcommand{\eee}{\end{equation}}
\newcommand{\bea}{\begin{eqnarray}}
\newcommand{\eea}{\end{eqnarray}}
\newcommand{\lb}[1]{\label{#1}}

\newcommand{\dd}{{\mathrm{d}}}
\newcommand{\up}[1]{\raisebox{0.7mm}{$\scriptstyle \, #1 $}}
\newcommand{\dn}[1]{\raisebox{-0.7mm}{$\scriptstyle #1 $}}
\usepackage{amsmath}
\usepackage{txfonts}
\usepackage[dvips]{graphicx}
\usepackage{epsfig}
\usepackage{color}
\usepackage{latexsym}
\usepackage{amsmath}
\usepackage{amssymb}
\usepackage{euscript}
\usepackage{pstricks}
\usepackage{picture}
\usepackage{empheq}
\def\[{\left\lbrack}
\def\]{\right\rbrack}

\def\({\left(}
\def\){\right)}

\def\ni{\noindent}

\begin{document}
\title{\Large Oscillations in the Tsallis income distribution}
\author{Everton M.\ C.\ Abreu}\email{Corresponding author:
        evertonabreu@ufrrj.br}
\affiliation{Departamento de F\'{i}sica, Universidade Federal Rural do
	     Rio de Janeiro--UFRRJ, Serop\'edica, RJ 23890-971, Brasil}
\affiliation{Departamento de F\'{i}sica, Universidade Federal de Juiz de
             Fora--UFJF, Juiz de Fora, Brasil}
\affiliation{Programa de P\'os-Gradua\c{c}\~ao Interdisciplinar em
	     F\'{\i}sica Aplicada, Instituto de F\'{\i}sica, Universidade
	     Federal do Rio de Janeiro-UFRJ, 21941-972, Rio de Janeiro,
             RJ, Brazil}
\author{Newton J. Moura Jr.}
\affiliation{Instituto Brasileiro de Geografia e Estat\' istica--IBGE,
	     Rio de Janeiro, Brasil}
\author{Abner D. Soares}
\affiliation{Comiss\~ao Nacional de Energia Nuclear--CNEN, Rio de Janeiro,
             Brasil}
\author{Marcelo B. Ribeiro}
\affiliation{Instituto de F\' isica, Universidade Federal do Rio de
             Janeiro--UFRJ, Rio de Janeiro, Brasil}
\affiliation{Observat\'orio do Valongo, Universidade Federal do Rio de
             Janeiro--UFRJ, Rio de Janeiro, Brasil}
\date{\today}
\begin{abstract}
\noindent Oscillations in the complementary cumulative distribution
function (CCDF) of individual income data have been found in the data
of various countries studied by different authors at different time
periods, but the dynamical origins of this behavior are currently
unknown. Although these datasets can be fitted by different functions
at different income ranges, the Tsallis distribution has recently been
found capable of fitting the whole distribution by means of only two
parameters. This procedure showed clearly such oscillatory feature in
the entire income range feature, but made it particularly visible at
the tail of the distribution. Although log-periodic functions fitted
to the data are capable of describing this behavior, a different
approach to naturally disclose such oscillatory characteristics is to
allow the Tsallis $q$-parameter to become complex. In this paper we
use this idea in order to describe the behavior of the CCDF of the
Brazilian personal income recently studied empirically by Soares et
al.\ (2016). Typical elements of periodic motion, such as amplitude
and angular frequency coupled to this income analysis, were obtained
by means of this approach. A highly non-linear function for the CCDF
was obtained through this methodology and a numerical test showed it
capable of recovering the main oscillatory feature of the original
CCDF of the personal income data of Brazil.
\end{abstract}
\maketitle

\section{Introduction}

The study of the individual income distribution of populations has a long
history. Vilfredo Pareto (1848-1927), the pioneer of this type of analysis,
studied the distribution of personal income at the end of 19th century
for some regions and countries in specific years and sets of years. Pareto
looked at the problem systematically and concluded that, individually
speaking, the richest people in a society have the \textit{complementary
cumulative distribution function} (CCDF) of income obeying a power law
function \cite{pareto}. Consequently, the \textit{probability density
function} (PDF) $p(x)$ of the personal income $x$ of the richest persons
may be given by,
\bee
p(x) = \beta\,x^{-(1+\alpha)}\,\,,
\label{paret}
\eee
where $\beta$ is a normalization constant. Through the years, this power
law behavior has become known as the \textit{Pareto power-law} and,
consequently, the exponent $\alpha$ is now known as the \textit{Pareto
index.} This law has been interpreted later as being a classic example
of a fractal distributions, where the Pareto index plays the role of the
single fractal dimension of the distribution \cite{mandelbrot}. Higher
values of the Pareto index imply in less uneven distribution of the
personal income. In other words, a rise in the Pareto index corresponds
to a fall in income inequality. The interesting detail is that this
result, that is, the power-law nature of the income distribution of the
richest persons in a society, has not been disputed by different
investigations carried out since then, which considered several different
samples obtained at different times for different populations in distinct
countries or groups of countries \cite[and references therein]{k80,nm09,
fnm10,dy01,ferrero2,s05,yr09,crbh08,by10,nm13,cyc05}.

Despite its empirical success, the Pareto power-law does not work for
the overwhelmingly majority less rich part of the population. Namely, it
only describes well the income data of those belonging to the narrow
``slice'' of the richest population. To consider the income data of the
group composed by the less rich people the method that has been used
since shortly after Pareto's time is to fit the less rich data segment
by various other functions, like the exponential, the log-normal, the
gamma function, the Gompertz curve, as well as other ones \cite{k80,nm09,
fnm10,ferrero2,nm13,cyc05,c5}. There are also successful approaches for
analyzing the whole data range using less than simple functions with many
parameters, but usually such approaches require four or more parameters
in order to fit the entire distribution. In addition, using the
two-fitting-functions methodology means assuming that societies are
divided in two classes only: on one side the very rich, formed by about
1\% of the whole population, and on the other side the remaining 99\%.
One basic problem of this methodology is the absence of a middle class,
which certainly exists in between these two groups. In addition, the
question remains of whether or not this class division is a real feature
of societies, or basically a mathematical artifact convenient for data
fitting. Considering these objections, a simple function able to fit the
entire income distribution which, at the same time, allows for various
features to emerge at different income range is certainly of high interest.

A recent approach for representing the whole income data is to fit the
data using the Tsallis functions instead of a combination of two functions
as depicted above. In this approach the individual income distribution is
classified in terms of the well known Tsallis parameter, \textit{i.e.},
the $q$-parameter, and another normalization constant. Borges \cite{borges}
used two $q$-parameter, where one controls the slope of the intermediate
income range and the other describes the tail of the distribution. He was
then able to analyze the income distribution concerning some \textit{counties}
of the USA from 1970 to 2000, Brazil from 1970 to 1996, Germany from 1992
to 1998 and the United Kingdom from 1993 to 1998. The conclusion was that
an increase in $q$ with time points to growing inequality. Greater values
of $q$ indicate greater probability to find counties much richer than
others. Ferrero \cite{ferrero2011,ferrero1} used the Tsallis function to
fit the entire income data of several \textit{countries}, but only at
single years, not being able though to indicate an evolution in the
Tsallis parameters.

Recently \cite{tsallis,tsallis-2} have used the Tsallis formalism to
empirically study the income distribution of Brazil during a relatively
large yearly time window, from 1978 to 2014. The results showed that the
two fitted parameters of the Tsallis function follow a cycling behavior
over time. Moreover, a linear fit of the distribution in each year showed
that the data oscillate periodically around the fitted straight line with
an amplitude that grows with the income values. A closer look at the
fitted data made by other authors using different methods applied to
different samples collected at different time periods showed a similar
oscillatory pattern, which means that there seems to be indeed a second
order dynamical effect not previously identified in the income data
\cite[p.\ 164]{smr}. This kind of oscillatory behavior has not been
noted before in the income distribution data, although it has been
observed in financial markets and other systems \cite{smr}.

In this work we have analyzed this periodic oscillation through the
alternative approach of allowing the Tsallis parameter to become
complex, as suggested by Ref.\ \cite{ww}. Under this methodology the
$q$-parameter discloses such periodic behavior in the income
distribution curves, allowing us to define periodic elements that are
ordinary in physics, like the amplitude and angular frequency. Such
methodology is exemplified by a numerical example where the original
empirically obtained CCDF of Brazil for the year 2011 is recovered
once our approach is applied to the results.

This paper is organized as follows. Sect.\ 2 presents the Tsallis
functions and some of their properties required in our analysis. Sect.\
3 presents and discusses the complexification process and its influence
in the analysis of the individual income distribution. Sect.\ 4 presents
a numerical example of the procedure developed here, and Sect.\ 5 ends
the paper with our conclusions.

\section{Tsallis functions}

It is well known that the Tsallis thermostatistics \cite{tsallis1,tsallis2}
is based on both the $q$-logarithm and $q$-exponential functions, given by,
\be
\ln_q x \equiv \frac{x^{(1-q)}-1}{1-q},
\lb{qlog}
\ee
\be
{{\mathrm{e}}_q}^x \equiv {\bigglb[1+(1-q)x \biggrb]}^{1/(1-q)}.
\lb{qexp}
\ee
These functions are defined such that for $q=1$ both expressions
become the standard logarithm and exponential functions, namely,
${\mathrm{e}_1}^x={\mathrm{e}}^x$ and $\ln_1 x=\ln x$. Hence, the
Tsallis $q$-functions are in fact the usual exponential and logarithmic
expressions twisted in such a way as to be used in Tsallis' theory of
non-extensive statistical mechanics \cite{tsallis2}.

At this point it should be noted that there are other ways to deform
these two common functions viewing other application, such as the
personal income distribution. This is the case of  the
$\kappa$-generalized exponential, introduced by Ref.\ \cite{cgk07},
which can be used to fit the entire income data range in similar manner
as the Tsallis $q$-functions. This is especially significant because
both of them have the power-law and exponential as their limiting cases.
The interested reader can find more applications of the
$\kappa$-generalized function in Refs.\ \cite{cmgk08,cgk09,cgk12}.

From the definitions above it is clear that,
\be
{{\mathrm{e}}_q}^{(\, \ln_q x)}= \ln_q \Big({\mathrm{e}_q}^x \,\Big)=x.
\lb{recip}
\ee
Moreover, $\ln_q 1=0$ for any $q$. Hence, if we have a value $x_0$ such
that $x/x_0=1$, as a result $\ln_q\,(x/x_0)=0$. Two other properties of
the $q$-exponential useful for our purposes here are as follows
\cite{yamano},
\be
{\left[ {\mathrm{e}_q}^{f(x)} \right] }^{\up{a}}=
{\mathrm{e}}^{\up{a f(x)}}_{\dn{\dn{{1-(1-q)/a}}}} \, ,
\lb{qexpa}
\ee
\be
\frac{\dd}{\dd x} \left[ {\mathrm{e}_q}^{f(x)} \right]=
\dfrac{{\left[ {\mathrm{e}_q}^{f(x)} \right]
}^{q}}{f {\displaystyle \,'}(x)}.
\lb{dfeqfx}
\ee

These results will be useful when we explore the fact that the Tsallis
parameter $q$ can be represented in the complex plane, as discussed in
Ref.\ \cite{ww}. It is worth noting that from Eqs.\ \eqref{qexp},
\eqref{qexpa} and \eqref{dfeqfx} it is not obvious that a complex $q$
can help us disclose some income distribution details that are hidden
in these functions, such as a periodic behavior, as we shall show below.

\section{$\mathbf{q}$-parameter complexification}

\subsection{Complex heat capacity}

The fact that the nonextensivity parameter $q$ can be represented by a
complex formulation is not new. In Ref.\ \cite{bbv} the $q$-parameter
can be seen as a measure of the thermal bath heat capacity $C$, where 
\bee
\label{q-C}
C=\frac{1}{q-1}\,\,.
\eee
Moreover, such complex $C$ is well known in the literature \cite{sg,sq-2}
and can be written as follows, 
\bee
\label{capacity}
C\,=\,C_\infty \,+\,\frac{C_0-C_\infty}{1+(\omega \tau)^2}\,
(1\,-\,\iu\omega\tau)\,\,,
\eee 
where $C_\infty$ is the heat capacity for infinitely fast degrees of
freedom (DOF), $\omega$ is the frequency, $C_0$ is the heat capacity
at equilibrium of DOF where the frequency is set to zero and $\tau$,
the time constant, is the kinetic relaxation time constant of a
certain DOF. The form of Eq.\ \eqref{capacity} suggests us that we
can write it as $C=C'\,+\,\iu\,C''$, where
\bee
C'=C_\infty \,+\,\frac{C_0-C_\infty}{1+(\omega\tau)^2}
\label{8.1}
\eee
and 
\bee
C''=\frac{(C_0-C_\infty)\omega\tau}{1+(\omega\tau)^2}\,\,.
\label{9.1}
\eee
Hence, let us write a complex form of $q$ as $q=q_r\,+\,\iu q_i$.
Substituting in Eq.\ \eqref{q-C} and associating with Eq.\
Eq.\ \eqref{capacity}, we have
that
\bee
\frac{q_r -1}{q_i}\,=\,\frac{C_0+C_\infty (\omega\tau)^2}
{(C_0-C_\infty)\omega\tau}\,\,,
\label{10.1}
\eee
and for $C_0<<C_\infty$ we have $(1-q_r)/q_i\,\simeq \omega\tau$,
which shows that the relationship between the real and imaginary
parts of $q$ is proportional to the frequency. More details of
this discussion can be found in Ref.\ \cite{bbv}.

\subsection{Complex income distribution means periodic behavior}

Let us start with the suggestion of Wilk and W{\l}odarczyk \cite{ww}
for the complexification of the $q$-parameter. A three-parameters
Tsallis distribution (TD) may be written as follows,
\bee
\label{1}
f(x) =C\,\mathrm{e}_q^{-x/T}= C\,\bigglb( 1-\frac{x}{mT}
\biggrb)^{-m}\,\,,
\eee
where
\bee
m=\frac{1}{(q-1)},
\label{mm}
\eee
is a real power index, $T$ is a scale parameter identified in
thermodynamic applications, in general the standard temperature,
and $C$ is a normalization constant. The proposal is to consider
$m$, or $q$, complex. Hence, the TD keeps its main quasi-power like
form, however, that brings about some log-periodic oscillations. As
examples, one can mention that such behavior has been encountered in
many subjects, such as earthquakes \cite{3,3-2}, chaos \cite{4},
tracers on random systems \cite{5,5-2,5-3}, random quenched and
fractals \cite{6,6-2,6-3,6-4,8}, specific heat \cite{7}, clusters
\cite{9}, growth models \cite{10}, stock markets \cite{11,11-1,11-2,
11-3,11-4} and, finally, non-extensive statistical mechanics
log-periodic oscillations \cite{12}. When $m \rightarrow \infty$,
namely, $q \rightarrow 1$, we have that this power-like distribution
is analogous to the standard exponential distribution
$f(x)=C\mathrm{e}^{-x/T}$.

The complexification proposal means turning $m$ complex in Eq.\
\eqref{1}, yielding,  
\bee
\label{2.0}
m=m'+\iu m''
\eee
which means that we can also have a complex non-extensive $q$-parameter
written as below,
\bee
\label{3}
q=1\,+\,\frac 1m = q' + \iu q''.
\eee
It is simple to see that
\bee
\label{4}
q'=1+\frac{m'}{|m|^2} \qquad \mbox{and} \qquad q'' = -\frac{m''}{|m|^2}
\eee
where
\bee
|m|^2\,=\,{m'}^2\,+\, {m''}^2.
\label{mmod}
\eee

Hence, the goal here is to analyze the results obtained in Ref.\
\cite{smr}, where the personal income distribution of Brazil shows a
periodic behavior as a function of the income variable for each yearly
sample, in the light of the complexification of the $q$-parameter.
Therefore, we wish to describe mathematically such an oscillatory
behavior.

\subsection{Non-extensive analysis of the income distribution of Brazil}

Let us now turn our attention to the main issue of this article. As
discussed in Ref.\ \cite{smr}, if the entire income distribution range
can be fitted by one function with only two parameters, a well-defined
two-classes-base income structure implicitly assumed when the income
range is described by two distinct functions may be open question.
Therefore, such income-class division could possibly be only a result
of fitting choices and not of an intrinsic feature of societies. The
TD is known to become a pure power-law for large values of its
independent variable $x$, and an exponential when $x$ tends to zero.
However, this behavior is not equivalent to assuming from the start a
two-classes approach to the income distribution problem because the TD
will only have power-law and exponential like behaviors as limiting
cases. Thus, a possible different behavior at the intermediate level
might not be described by neither of these functions. This means that
the TD does not necessarily imply in two very distinct classes based on
well-defined income domain ranges, but possibly having an intermediate
income range of unknown size which might behave as neither of them.
Bearing these points in mind, let us now proceed with the description
of the income distribution in terms of the TD and its subsequent
complexification.

Let  ${\cal F}(x)$ be the \textit{cumulative distribution function} (CDF)
of the personal income, representing the proportionality, or probability,
that a person receives an income less than or equal to $x$. Let us now
denote its complementary version, the CCDF, by $F(x)$, which then
describes the probability that a person receives an income greater or
equal to $x$. It is then clear that,
\bee
\label{5}
{\cal F}(x) + F(x) = 100\,\,.
\eee
Here the maximum probability is normalized to $100{\%}$ instead of the
standard unity value. The boundary conditions involved in both functions
are ${\cal F}(x)=F(\infty)\cong 0$ and ${\cal F}(\infty)=F(0) \cong 100$.
In addition, the following properties apply to these income functions,
\bee
\frac{\dd {\cal F}(x)}{\dd x}=-\frac{\dd F(x)}{\dd x}=f(x),
\label{fff}
\eee
\bee
\label{6}
\int^{\;\infty}_0 f(x)\,\dd x = 100 \,\,,
\eee
where $f(x)$ is the PDF \cite{nm13,fnm10}.

The empirical suggestion that the income distribution can be modeled
by the TD comes from the fact that when $F(x)$ is obtained from income
data and plotted in a log-log scale, its functional curve decreases as
the income $x$ increases. In addition, the general shape of the
empirical CCDF function, particularly its ``belly'', is analogous
to the behavior of $\mathrm{e}^{-x}_{q}$ for $q > 1$ when plotted in a
log-log scale \citetext{see Ref.\ \citealp{tsallis2}, p.\ 40, Fig.\ 3.4}.
Besides, as mentioned above the Tsallis functions have power-law like
behavior for high income values, agreeing then with the Pareto power-law.
Taking together these observations into account, Ref.\ \cite{smr} advanced
the following description for the individual income distribution, 
\bee
\label{7a}
F(x)=A\mathrm{e}^{-Bx}_q \,\,,
\eee
where $A$ and $B$ are positive parameters. Since we have a boundary
condition in the form of $F(0)=100$, the expressions above implies that
$A=100$. Hence, Eq.\ \eqref{7a} may be rewritten as follows,
\bee
\label{8}
F(x)=100\,\mathrm{e}^{-Bx}_q \,\,.
\eee
Note that a cursory examination of Eq.\ \eqref{8} does not present any
obvious evidence of a periodic behavior, or that the complexification
of $q$ will disclose any oscillatory feature, since the $q$-parameter
is just an index. We shall show below that allowing $q$ to become
complex will expose such features.

\subsection{Periodic behavior of income distribution function}

Let us start with the definition \eqref{qexp} in order to rewrite
Eq.\ \eqref{8} as below,
\bee
\label{9}
F(x)\,=\,100\,\Big[1\,+\,(1-q)(-Bx) \Big]^{1/(1-q)} \,\,.
\eee
This equation can also be expressed as a function of $m$ using Eq.\
\eqref{mm}, 
\bee
\label{10}
F(x)\,=\,100\, \bigglb[ 1\,+\, \frac Bm x \biggrb]^{-m}.
\eee
Following the complexification suggested in Eqs.\ \eqref{2.0} to
\eqref{mmod}, Eq.\ \eqref{10} may be written as below,
\bee
\label{D}
F(x)\,=\,100\,\Big[ a(x)\,+\,\iu b(x) \Big]^{-(m'+\iu m'')},
\eee
where
\bee
a(x)\,=\,1\,+\,\frac{B\,m'}{|m|^2}\,x\;,
\label{ax}
\eee
\bee
b(x)\,=\,-\,\frac{B\,m''}{|m|^2}\,x\,\,.
\label{bx}
\eee
Expanding the terms in Eq.\ \eqref{D} results in the following expression,
\bee
\label{final}
F(x)\,=100\,{\cal A}(x)\,\mathrm{e}^{-\iu\omega (x)}\,=100\,{\cal A}(x)\,
\Big[ \cos \omega(x) \,-\iu\sin \omega (x) \Big],
\eee
where
\begin{empheq}[left=\empheqlbrace]{align}
{\cal A}(x)&=\exp \left[m''\varphi(x)-m'\theta(x)\right],
\label{Ax} \\[0.25cm]
\omega(x)&=m'\varphi(x)+m''\theta(x),
\label{omegax}
\end{empheq}
and
\begin{empheq}[left=\empheqlbrace]{align}
\varphi(x)&=\tan^{-1} \left[\frac{-B\,m''x}
{\left(|m|^2+B\,m'x\right)}\right],
\label{phix} \\[0.30cm]
\theta(x)&=\ln\sqrt{1+\frac{Bx}{|m|^2}\left( Bx+2m' \right)}\;.
\label{tetax}
\end{empheq}
We have to remember that $B$ is always positive. Now, if we consider
only the real part of Eq.\ \eqref{final}, the original description of
the income data in terms of the TD, as given by Eq.\ \eqref{8}, turns
out to be written as follows, 
\bee
F(x)=100\,\mathrm{e}^{-Bx}_q=100\,{\cal A}(x)\,\Big|\cos\omega(x)
\Big|\,\,,
\label{final2}
\eee
where the absolute value of $\cos \omega(x)$ guarantees that we will
always have the empirically obtained positive values for $F(x)$.

This result shows clearly that the periodic behavior empirically
observed by Ref.\ \cite{smr} in the income data seems to be described
by the expression above. Or that the observed oscillatory behavior can
at least be expected, since it is built in the TD. Note that the original
$q$-parameter is present in both ${\cal A}(x)$ and ${\omega}(x)$ through
$m'$ and $m''$, which are the respective real and imaginary parts of $q$.

The fact that the individual income distribution function can be written
in the standard exponential form given by Eq.\ \eqref{final} shows us
that we can define a periodic function where ${\cal A}(x)$ can be seen
as the amplitude of this oscillatory motion and ${\omega}(x)$ its angular
frequency. Considering the $q$-parameter only by a real term hides this
periodic behavior. The complexification adds new components such as the
kind derived above so that it is able of revealing a periodic behavior
that does appear in the empirical curves obtained from the income
distribution data.

The issue of always considering the $q$ parameter as an entirely complex
number, namely with both real and imaginary parts, is an open question
although we believe that it depends on the problem we are dealing with.
We can imagine an analogy of this feature with the dual characteristic
of the electron in quantum physics, since it has a wave-particle
duality behavioral feature that depends on the experience that we are
analyzing. In this way we could talk about a $q$-duality, which would
deserves alternative interpretations and could  certainly be a target of
further research.

\section{Numerical application}

Our problem now is to analyze the $q$-complex version of Eq.\
(\ref{final2}) concerning income distribution real data through a
numerical example. The aim is to calculate the values of ${m}'$ and
${m}''$ from Eq.\ \eqref{2.0}, that is, to obtain for each empirical
income value $x_k$ given in the tables of Ref.\ \cite{smr} a pair
${m_k}'$ and ${m_k}''$.

Let us begin by rewriting Eqs.\ \eqref{phix} and \eqref{tetax} to
conform with discrete real numerical data, as follows,
\bee
\label{2.1}
\varphi_k \,=\,\varphi\,(x_k)\,=\,  \arctan\, 
\Bigglb(\frac{-B{m_k}'' x_k}{{{m_k}'}^2 + {{m_k}''}^2 + B{m_k}' x_k}
\Biggrb),
\eee
\bee
\label{3.teta}
\theta_k\,=\,\theta\, (x_k)\,=\,  
\ln\sqrt{1+ \frac{Bx_k \Big(B x_k + 2 {m_k}' \Big)}{{{m_k}'}^{2} +
{{m_k}''}^2}} \,\,,
\eee
where $x_k$ is the $k$-th income data point as produced by Ref.\
\cite{smr}. From Eq.\ \eqref{Ax} we can write the following expression
for the $k$-th tabled income data, 
\bee
\label{4b}
{\cal A}_k\,=\,{\cal A} (x_k)\,=\, \exp\,\left[\,{m_k}'' \, \arctan\, 
\Bigglb(\frac{-B{m_k}'' x_k}{{{m_k}'}^2 + {{m_k}''}^2 + B{m_k}' x_k} \Biggrb)
\,-\,{m_k}' \ln\sqrt{1+ \frac{Bx_k \Big(B x_k + 2 {m_k}'\Big)}{{{m_k}'}^{2} 
+ {{m_k}''}^2}} \,\right] \,\,.
\eee
Similarly, Eq.\ \eqref{omegax} allows us to write the expression below,
\bee
\label{5b}
\omega_k\,=\,\omega(x_k)\,=\,{m_k}'\arctan\Bigglb[\frac{-B{{m_k}''}x_k}
{{{m_k}'}^2\,+\,{{m_k}''}^2\, +\,B{m_k}' x_k}\Biggrb]\,+\,{m_k}'' \ln
\sqrt{1\,+\,\frac{B x_k \Big(B x_k + 2{m_k}'\Big)}{{{m_k}'}^2\,+\,
{{m_k}''}^2}} \,\,.
\eee

The aim is to solve Eq.\ \eqref{final2} numerically. To accomplish this
task, let us write it in the following numerically discrete form, 
\bee
\label{8.11}
F(x_k)\,=\,100\,{\cal A}(x_k)\, \Big|\cos \omega\,(x_k)\Big|, \qquad
\Longrightarrow \quad F_k\,=\,100\,{\cal A}_k\,\Big|\cos \omega_k \Big|. 
\eee
For the $k$-th income one needs to calculate both ${m_k}'$ and ${m_k}''$.
From Eqs.\ \eqref{3} to \eqref{mmod} we have that 
\bee
\label{8.1a}
|m|^2\,=\,\frac{1}{|q-1|^2}\,=\,{{m_k}'}^2\,+\,{{m_k}''}^2,
\eee
which allows us to write the expression below, 
\bee
\label{8.2}
{{m_k}''}\,=\, \pm\,\frac{\sqrt{1\,-\,{{m_k}'}^2\,|q-1|^2}}{q-1} \,\,.
\eee
Considering that there are two solutions for ${{m_k}''}$ due to the square
root, substituting Eq.\ \eqref{8.2} into Eqs.\ \eqref{4b} and \eqref{5b} we
respectively obtain the results below,
\bea
\label{7}
{\cal A}_k\,&=&\, \exp\,\Bigglb\{\,-\,\frac{\sqrt{1-{{m_k}'}^2 |q-1|^2}}
{q-1}\, \arctan\, \Bigglb[\frac{B x_k (q-1) \sqrt{1-{{m_k}'}^2 |q-1|^2}}
{1 + B{m_k}' x_k (q-1)^2} \Biggrb]- \nonumber \\
\,&-&\,{m_k}' \ln\sqrt{1+ B x_k (q-1)^2 \Big(B x_k + 2 {m_k}'\Big)} \,
\Biggrb\},
\eea
and
\bea
\label{7.2}
\omega_k\,&=&\, -\,\,{m_k}'  \arctan\, \Bigglb[\frac{B x_k (q-1) 
\sqrt{1-{{m_k}'}^2 |q-1|^2} }{1 + B{m_k}' x_k (q-1)^2} \Biggrb]+\nonumber \\
\,&+&\,\frac{\sqrt{1-{{m_k}'}^2 |q-1|^2}}{q-1}\ln\sqrt{1+ B x_k (q-1)^2 
\Big(B x_k + 2 {m_k}'\Big)}\;. 
\eea
Finally, substituting both expressions above into Eq.\ \eqref{8.11} the
result may be written as below,
\bea
\label{9b}
{F_k}\,&=&\,100\exp\,\Bigglb\{\,-\,\frac{\sqrt{1-{{m_k}'}^2 |q-1|^2}}{q-1}
\, \arctan\, \Bigglb[\frac{B x_k |q-1| \sqrt{1-{{m_k}'}^2|q-1|^2}}{1+
B{m_k}' x_k |q-1|^2} \Biggrb]- \nonumber \\
&-&\,{m_k}' \ln\sqrt{1+ B x_k |q-1|^2 \Big(B x_k + 2 {m_k}'\Big)} \,
\Biggrb\} \times \nonumber \\
&\times& \Bigglb|\cos\,\Bigglb\{\,-\,{m_k}'  \arctan\, \Bigglb[\frac{B x_k
|q-1| \sqrt{1-{{m_k}'}^2|q-1|^2}}{1+ B{m_k}' x_k |q-1|^2} \Biggrb]+
\nonumber \\
\,&+&\,\frac{\sqrt{1-{{m_k}'}^2 |q-1|^2}}{q-1} \ln\sqrt{1+ B x_k |q-1|^2 
\Big(B x_k + 2 {m_k}'\Big)} \,\Biggrb\} \Biggrb| \,\,.
\eea

Ref.\ \cite{smr} used income data of Brazil to obtain the distribution
$F_k$ in terms of income values $x_k$ for each year in the observed time 
window and then fitted the TD to the empirical distribution in order to
find both parameters $q$ and $B$ for all $k$-values in a given year. Once
this was done, knowing the values of $[q,B]$ in that given year one can
solve Eq.\ \eqref{9b} numerically for each data pair $[F_k,x_k]$ to finally
obtain the unknown quantity ${m_k}'$.

The methodology described above can then be used to test the whole
complexification procedure. One starts by choosing the dataset of a single
year as provided by Ref.\ \cite{smr}, calculate ${m_k}'$ for each data
point using the empirical values $[F_k,x_k,q,B]$ in the chosen year to
find the roots of the expression \eqref{9b}, substitute the results 
$[{m_k}',x_k,q,B]$ back into Eq.\ \eqref{9b} to recover the distribution
$F_k$ and then compare the original distribution with the recovered one.
If the recovered points follow closely the original empirical distribution,
including the observed oscillation, that would demonstrate that the 
complexification procedure above really discloses the oscillations present
in the distribution. 

Fig.\ \ref{plot} shows the results obtained with this approach using the
empirical CCDF for the income data of Brazil in the year 2011. One can
\begin{figure}[htbp]
\begin{center}
\hspace{2cm}\includegraphics[trim=-0.2cm 0 0 0,scale=1.6]{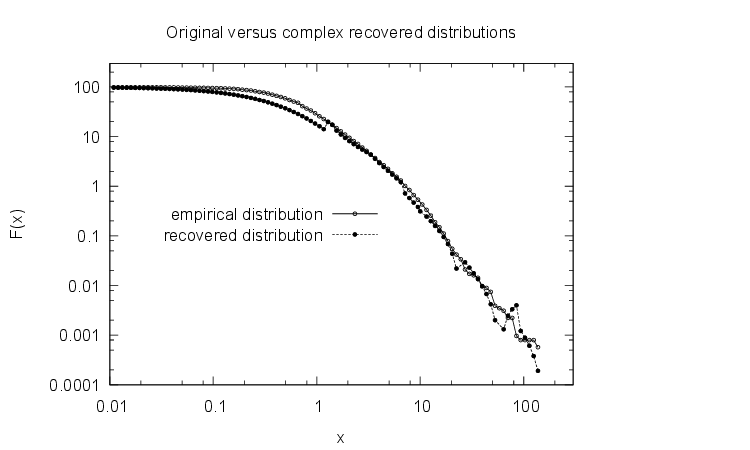}
\end{center}
\caption{This plot presents the CCDF of Brazil for the year 2011 with
	$q=1.265$ obtained with the empirical data (open circles) and the
	recovered ones (filled circles) obtained by means of finding the
	roots of Eq.\ \eqref{9b}. Eqs.\ (\ref{1}) and (\ref{mm})
	are the parameter transformation from $q$ to $m$, and although $q$
	has a single value for the whole distribution, the complexification
	process produces real and imaginary parts \textit{for each point of
	the distribution}. So, as the distribution is made of $k$ points
	there will be $k$ values for complex $m$, as shown by Eqs.\
	(\ref{8.1a}), (\ref{8.2}) and (\ref{9b}). Therefore, the $k$ points
	produced by the solution of Eq.\ (\ref{9b}) are shown in this plot.
	Since Eq.\ (\ref{9b}) is actually an expanded form of Eq.\
	(\ref{8.11}), and that it has an oscillatory term coupled with an
	amplitude term, then, once the $k$ values of $F_k$ are found by
	Eq.\ (\ref{9b}), the oscillatory nature of the distribution is
	reproduced by the fitting. However, due to its high non-linearity
	and the fact that it has several subtractions of two values close
	to unit, the numerical problem suffered from strong numerical
	instability due to the catastrophic loss of significant digits.
	This instability was significantly ameliorated by the use of 20
	digits during the numerical evaluations of Eq.\ \eqref{9b}.
	Despite the instability, the oscillatory nature of the distribution
	is clearly visible in the recovered points, meaning that allowing
	the parameter $q$ of the TD to become complex results in revealing
	the oscillatory nature of the distribution, especially at its
	tail.}
\lb{plot}
\end{figure}
clearly see that the recovered distribution does follow closely the
empirical one, including the more pronounced oscillation at the tail due
to increasing values of the amplitude ${\cal A}_k$. These results could
not bettered due to high non-linearity of Eq.\ \eqref{9b}, since calculating
the root of this expression resulted in strong numerical fluctuations
due to catastrophic loss of significant digits. Reducing these fluctuations
required the use of no less than 20 digits in the numerical evaluation. Using
more than 20 digits did not improve the results because the parameters $[q,B]$
were fitted to the data with up to 3 digits only.

The results shown in Fig.\ \ref{plot} could, perhaps, be improved if the
numerical fluctuations due to the loss of significant digits were to be
somehow reduced. One possibility for doing that would be by means of
recalculating both parameters $q$ and $B$ with at least 20 digits, since
they were both originally obtained with only 3 digits. But, the task of
reducing these fluctuations is beyond the scope of this paper because our
aim here is just to show that allowing the parameter $q$ of the TD to become
complex results in revealing the oscillatory nature of the distribution, as
demonstrated by the graphs of Fig.\ \ref{plot}.

\section{Conclusions}

Since the Pareto's work the study of the income distribution of the whole
population has been a target of economic experts, and for some time
now, to the econophysics literature. Despite the relative great number of
parameters used in several functions to fit the income data, from three to
five in some cases, some mathematical approaches have thrived in the
description of the entire data range. In this work we have used Tsallis'
non-extensive point of view with a complexification mode where its 
$q$-parameter is represented by a complex number. The objective here was
to use this complexification in order to justify analytically the results
obtained in Ref.\ \cite{smr}, which show a periodic behavior in the income
data.

As shown above, in doing this we, however, increased the number of
parameters required to fit the data, from the original two to three.
Although increasing the number of unknown parameters in a problem is a
bad procedure, in our case the complexification has disclosed an extra
behavior of the income distribution in the form of a periodic motion in
the income distribution, motion which was already present in several
studies of income distribution, but was only explicitly acknowledged by
Soares {\it et al.\ } \cite{smr}. By interpreting this oscillatory
motion of the income distribution as a typical periodic motion allowed
us to define commonly used oscillatory parameters such as amplitude and
angular frequency.

The analytical procedure developed here was tested against real data,
in this case the income distribution data of Brazil in 2011, and the
numerical results showed that allowing for a complex $q$ parameter
results in revealing the oscillatory nature of the distribution,
especially at its tail. So, from the results obtained here we can say
that ``the highs and lows'' of the yearly income distribution samples
are an expected behavior, confirmed using the complex form of the
Tsallis $q$-parameter. However, it not all clear that this periodic
oscillation could lead us to the understanding of the real nature of
the $q$-parameter, namely, if it is complex or not depends on the
features of the problem we are dealing with.

Finally, this work shows us a glimpse of the task ahead as far as the
empirical studies of income distribution are concerned. The data give
us $F(x), x, B, q$ and $m$, so the empirical task is to determine both
$m'$ and $m''$ from the data in order to end up with only three parameters
required to characterize the whole yearly distribution, oscillatory
feature included, namely $B$, $m'$ and $m''$. 

\section{Acknowledgments}

\ni E.M.C.A. thanks the Brazilian Federal Agency CNPq Conselho Nacional de 
Desenvolvimento Cient\' ifico e Tecnol\'ogico), a federal scientific support 
agency, for partial financial support, Grants numbers 302155/2015-5 and 406894/2018-3. 
M.B.R. acknowledges partial financial support from the Rio de Janeiro State 
Scientific Funding Agency (FAPERJ).


\end{document}